\documentclass[preprint,prd,aps]{revtex4-1}

\usepackage{amsmath,amssymb,amsfonts,dcolumn,color,graphicx,graphics,latexsym,placeins,epsfig}
\usepackage{subfigure}
\usepackage{array}
\newcommand{\be}{\begin{equation}}
\newcommand{\ee}{\end{equation}}
\newcommand{\ba}{\begin{eqnarray}}
\newcommand{\ea}{\end{eqnarray}}

\newcommand{\rr}{\raggedright}
\newcommand{\tn}{\tabularnewline}

\begin{document}

\title{\Large \bf Cosmology in a reduced Born-Infeld--$f(T)$ theory of gravity}

\author{Soumya Jana}
\email{soumyajana@phy.iitkgp.ernet.in}
\affiliation{\rm Department of Physics {\it and} Center for Theoretical Studies \\Indian Institute of Technology, Kharagpur, 721302, India}

\begin{abstract}
A perfect fluid, spatially flat cosmology in a $f(T)$ model, derived from a recently proposed general Born-Infeld type theory of gravity is studied. Four dimensional cosmological solutions are obtained assuming the equation of state $p=\omega \rho$.  For a positive value of $\lambda $ (a parameter in the theory) the solution is singular (of big-bang type) but may have accelerated expansion at an early stage. For $\lambda<0$ there exists a non-zero minimum scale factor and a finite maximum value of the energy density, but the curvature scalar diverges. Interestingly, for $\lambda <0$, the universe may undergo an eternal accelerated expansion with a de Sitter expansion phase at late times. We find these features without considering any extra matter field or even negative pressure. Fitting our model with Supernova data we find that the simplest dust model ($p=0$), with $\lambda >0$, is able to generate acceleration and fits well, although the resulting properties of the universe differ much from the known, present day, accepted values. The best fit model requires (with $\lambda > 0$) an additional component of the physical matter density, with a negative value of the equation of state parameter, along with dust. The $\lambda < 0$ solutions do not fit well with observations. Though these models  do not explain the dark energy problem with consistency, their analysis does shed light on the plausibility of an alternative geometrical explanation.
\end{abstract}


\maketitle

\section{\bf Introduction} 

\noindent Though general relativity (GR) remains so far the most successful classical theory of gravity, it has been under scrutiny for a long time, particularly in the strong field regime. Many long standing puzzles such as the resolution of singularities in GR, explanation of the dark energy problem have led researchers to pursue alternative/modified theories of gravity in the classical framework and also in quantum theory as well. One such modification is inspired by the well known Born-Infeld electrodynamics where we are able to regularize the infinity in the electric field at the location of a point charge \cite{born}. With a similar determinantal structure $\left[\sqrt{-det(g_{\mu\nu}+bR_{\mu\nu})} \right]$ as in the action of Born-Infeld electrodynamics, a gravity theory in the metric formulation was suggested by Deser and Gibbons \cite{desgib}. In fact, the determinantal form of the gravitational action existed much before through Eddington's formulation of GR in de Sitter spacetime \cite{edd}. This formulation is affine and the connection is the basic variable instead of the metric. However, coupling of matter remained a problem in Eddington's approach.

\noindent A pure metric formulation of the Deser-Gibbons proposal may give rise to a higher derivative theory of gravity which generally suffers from the ghost instability problem \cite{ghost}. One way to get rid of this is to use a Palatini formulation \cite{vollick}, where the metric and connection are treated independently. As a result, one gets field equations which do not contain fourth or higher order derivatives of the metric. Along this line of thought, recently, Banados and Ferreira have come up with a modified gravity theory which is now popularly known as Eddington-inspired Born-Infeld (EiBI) gravity \cite{banados}. Interestingly  this theory has non-singular cosmological solutions \cite{banados,eibicosmo}. Various astrophysical \cite{eibiastro} and other \cite{eibiother} aspects have been studied in  the context of this theory, by different authors.

\noindent Another approach to obtain a second order theory is to use the teleparallel formulation, where the basic variable is the {\em vierbein} or {\em tetrad} of the local Lorentz frame instead of the metric or the connection. The gravitational action contains first order derivatives of the {\em tetrad} field. The idea of teleparallelism originally belongs to Einstein in his attempt to unify gravitational and electrodynamic interaction through a single quantity called the {\em torsion tensor} \cite{einstein}. Though the attempt was unsuccessful, it gave an alternative description where the gravitational interaction arises through torsion, instead of curvature. Such description is known as the teleparallel equivalent of GR (TEGR). It is important  because it allows us to interpret gravity as a gauge theory \cite{aldrovandi}. Recently a new class of modified theories of gravity have been proposed where the {\em torsion scalar}, $T$ in the TEGR Lagrangian, is replaced by a general $f(T)$, which may explain the accelerated expansion of the universe \cite{bengochea,linder}. Cosmological solutions have been worked out with different forms of $f(T)$ by various authors \cite{f(t)cosmo}.

\noindent In this paper, we consider a $f(T)$ model reduced from the general Born-Infeld type theory of gravity proposed by Fiorini \cite{fiorini}. We work out cosmological solutions in the $3+1$ dimensional model. We have found some new results which are different from those shown in previous Born-Infeld type models \cite{fiorini,ferraro,banados, eibicosmo}. Our $3+1$ solution is an analytical one. We also test the model observationally by fitting the supernova data. We have organized our paper in the following three sections. In Section \ref{sec:theory} we briefly outline the $f(T)$ model on which we work. In Section \ref{sec:cosmology} we show an application in cosmology. Finally, in Section \ref{sec:conclusion}, we summarize our results and conclude.         

\section{Reduced $f(T)$ theory from a Born-Infeld theory of gravity }
\label{sec:theory}
\noindent We first give a brief outline of the $f(T)$ theory of gravity \cite{bengochea} which is a generalisation of the teleparallel equivalent of General Relativity (GR). In teleparallelism \cite{einstein,aldrovandi}, the fundamental field is not the {\em metric} tensor, instead it is the {\em vierbein} or $tetrad$, ${\bf e}_{i}$ $(x^{\mu})$ defining an orthonormal basis for the tangent space at each spacetime point ($x^{\mu}$) on the manifold. Here Latin indices ($i,j,k...$) and Greek indices ($\mu,\nu,\rho....$) refer respectively to a local Lorentz frame in the tangent space  and general spacetime coordinates of the manifold. The tetrad field ${\bf e}_i$ (or its dual ${\bf e}^i$) can be decomposed into {\em vector} components $e_{i}^{\; \mu}$ (or inversely $e^{i}_{\; \mu}$) such that: ${\bf e}_{i}\equiv \frac{\partial}{\partial\xi^i}$ $= e_{i}^{\; \mu}\partial_{\mu}$ or  ${\bf e}^{i}\equiv d\xi^{i}$ $=e^{i}_{\; \mu}dx^{\mu}$ (where, $\xi^i$ is the coordinate in the local Lorentz frame) and $e_{i}^{\; \mu}e^{i}_{\; \nu}=\delta^{\mu}_{\nu}$, $e_{i}^{\; \mu}e^{j}_{\; \mu}=\delta^{j}_{i}$. The distance function, $ds^2=\eta_{ij}{\bf e}^i {\bf e}^{j}=g_{\mu\nu}dx^{\mu}dx^{\nu}$. So, the relations between the {\em metric} tensor components and the {\em tetrad} field components are
\begin{equation}
g_{\mu\nu}(x)=\eta_{ij}e^{i}_{\; \mu}(x)e^{j}_{\; \nu}(x)
\label{eq:metric}
\end{equation}  
where $\eta_{ij}=diag.(-1,1,1,1)$ (in 4D Minkowski space). Instead of using the Levi-Civita connection in Riemannian geometry, here, in absolute parallelism, the curvature-less Weitzenb\"{o}ck connection \cite{weitzenbock} is used, in an attempt to encode the gravitational effects in torsion. The Weitzenb\"{o}ck connection is $\tilde{\Gamma}^{\rho}_{\; \mu\nu}=e_{a}^{\; \rho}\partial_{\nu}e^{a}_{\; \mu}$ and thus, we construct the torsion tensor whose components are
\begin{equation}
T^{\rho}_{\; \mu \nu}= \tilde{\Gamma}^{\rho}_{\; \nu \mu}- \tilde{\Gamma}^{\rho}_{\; \mu \nu}=e_{a}^{\; \rho}\left(\partial_{\mu}e^{a}_{\; \nu}-\partial_{\nu}e^{a}_{\; \mu}\right)
\label{eq:torsion}
\end{equation}
The Lagrangian density for teleparallel equivalent action of general relativity (TEGR) \cite{hayashi,maluf} is given by
\begin{equation}
\mathcal{L}_T\equiv \frac{e}{16\pi G} T =\frac{e}{16\pi G} S_{\rho}^{\; \mu\nu}T^{\rho}_{\; \mu\nu}
\label{eq:tegr}
\end{equation}
where $T$ is called as  {\em Torsion scalar}, $e$ is determinant of the matrix $e^{a}_{\; \mu}$ (one can show $e=\sqrt{-\vert g_{\mu\nu}\vert}$, where $\vert g_{\mu\nu}\vert$ is the determinant of metric tensor) and $S_{\rho}^{\; \mu\nu}$ is defined as
\begin{equation}
S_{\rho}^{\; \mu\nu}=\frac{1}{4}\left(T_{\rho}^{\; \mu\nu}-T^{\mu\nu}_{\; \; \; \rho}+T^{\nu\mu}_{\; \; \; \rho}\right)+\frac{1}{2}\delta^{\nu}_{\rho}T_{\sigma}^{\; \sigma \mu}-\frac{1}{2}\delta^{\mu}_{\rho}T_{\sigma}^{\; \sigma \nu}
\label{eq:stensor}
\end{equation} 
Such a {\em torsion scalar} differs from the {\em Ricci scalar} made up of Levi-Civita connection, by a total derivative \cite{maluf}. Hence the TEGR action [Eq.~(\ref{eq:tegr})] is completely equivalent to the Einstein-Hilbert action and reproduces GR. To have a modified theory of gravity, one can generalise the TEGR action replacing $T$ by a general functional $f(T)$ in Eq.~(\ref{eq:tegr}). The variation of this modified action, including the matter part to it, with respect to {\em vierbein} ($e_{a}^{\; \mu}$) leads to the field equation \cite{bengochea} (see also \cite{f(t)fieldeq} for a detailed derivation)
\begin{equation}
f_{TT}(T)S_{\mu}^{\; \rho \nu}\partial_{\rho}T+f_T(T)\left[\frac{1}{e}e^{a}_{\; \mu}\partial_{\rho}\left(ee_{a}^{\; \beta}S_{\beta}^{\; \rho \nu}\right)+T^{\rho}_{\; \beta\mu}S_{\rho}^{\; \beta\nu}\right]-\frac{1}{4}\delta^{\nu}_{\mu}f(T)=-4\pi G\Theta^{\; \nu}_{ \mu}
\label{eq:f(t)_fieldeq}
\end{equation}    
where $f_{TT}=\frac{d^2f}{dT^2}$, $f_{T}=\frac{df}{dT}$ and $\Theta^{\; \nu}_{\mu}$ is the usual energy-momentum tensor coupled to the {\em metric}. The field equation (\ref{eq:f(t)_fieldeq}) can further be rewritten in a more convenient form \cite{barrowf(t)eq}
\begin{equation}
f_TG_{\mu\nu}+\frac{1}{2}g_{\mu\nu}\left[f-Tf_T\right]+B_{\mu\nu}f_{TT}=8\pi G\Theta_{\mu\nu}
\label{eq:f(t)_covariant}
\end{equation}
where $B_{\mu\nu}=2S_{\nu\mu}^{\; \; \; \alpha}\nabla_{\alpha}T$ and $G_{\mu\nu}$ are components of the {\em Einstein tensor}. Note that if $f(T)\equiv T+const.$, then Eq.~(\ref{eq:f(t)_covariant}) reduces to the Einstein's field equations in GR. Another important point is that $f(T)$ theory does not respect {\em local Lorentz invariance} in contrast to GR or TEGR \cite{barrow_f(t)lli}.
       
\noindent The teleparallel equivalent
of Born-Infeld type theories has also been pursued by some authors \cite{ferraro,tebi}. Fiorini \cite{fiorini} has proposed a 
theory where a very interesting result of a non-singular early universe with a natural inflationary phase without any inflationary field has been found. Fiorini, in his formalism, exploited the same generally covariant 
determinantal structure introduced by Born and Infeld \cite{born}.  
 The action for the Born-Infeld type gravity, proposed in \cite{fiorini} is 
\begin{equation}
I_{BIG}=\frac{\lambda}{16\pi G}\int d^{D}x\left[\sqrt{-\vert g_{\mu\nu}+\frac{2}{\lambda}F_{\mu\nu} \vert}-\sqrt{-\vert g_{\mu\nu} \vert}\right]
\label{eq:fiorini}
\end{equation}
where $F_{\mu\nu}=\alpha S_{\mu}^{\; \sigma \rho}T_{\nu\sigma\rho}+\beta S^{\; \; \;\rho}_{ \sigma \mu}T^{\sigma}_{\;\nu\rho}+\gamma g_{\mu\nu}T$. In (\ref{eq:fiorini}), $\alpha,\, \beta,\, \gamma $ are arbitrary constant parameters satisfying a constraint relation $\alpha +\beta + \gamma D =1$, where $D$ is the dimensionality of the theory. The trace of $F_{\mu\nu}$ equals the {\em torsion scalar} and in the low energy limit ($\lambda \rightarrow \infty$) the action reduces to that of TEGR (\ref{eq:tegr}). Various combinations of $\alpha,\, \beta,\, \gamma $ are considered. The inflationary phases of the early universe in both spatially flat and curved cosmology are explored in \cite{fiorini} and the occurrence of different types of cosmological singularities are noted in \cite{suddensingularity}. In our case here, we consider $\alpha=\beta=0$. Under this assumption, the action (\ref{eq:fiorini}) reduces to a $f(T)$ type action 
\begin{equation}
I_{BI0}=\frac{\lambda}{16\pi G}\int d^Dx\, e\left[ \left(1+\frac{2T}{D\lambda}\right)^{D/2}-1\right]
\label{eq:bif(t)}
\end{equation}  
The form of $f(T)$ is identified as
\begin{equation}
f(T)=\lambda \left[ \left(1+\frac{2T}{D\lambda}\right)^{D/2}-1\right]
\label{eq:f(t)form}
\end{equation}
In \cite{fiorini} this action is given but its consequences have not been studied in detail. We obtain various interesting features of solutions, in the following section. 

\section{Cosmology}
\label{sec:cosmology}
\noindent We work out the cosmology in a $4D$ model ($i.e.$ for $D=4$ in Eqs.(~\ref{eq:bif(t)}),(~\ref{eq:f(t)form})). A homogeneous and isotropic FRW spacetime which is spatially flat, in $D$-dimensions has a line element given as 
\begin{equation}
ds^2=-dt^2+a^2(t)\left[dx_1^2+dx_2^2+.....+dx_{D-1}^2\right]
\label{eq:frwspacetime}
\end{equation}
We assume the matter sector to be that of a perfect fluid having energy-momentum tensor, $\Theta_{\mu\nu}=\left(p+\rho\right)u_{\mu}u_{\nu}+pg_{\mu\nu}$, where  $\rho$ is the energy density and $p$ is the pressure. We first write the field equations in $D$-dimensions and then look at the $D=4$ case. We also summarize the results for the $D=3$ case as a toy model, in the Appendix.  We choose $vierbein$ field as diagonal, $\lbrace e^{a}_{\; \mu}\rbrace=diag.\lbrace 1,a(t),a(t),....\rbrace$. Using previously stated relations we write down the {\em torsion scalar}, the non-zero $B_{\mu\nu}$, non-zero components of the {\em Einstein tensor}, $ f_T(T),\, f_{TT}(T)$ as
\begin{eqnarray}
T&=&(D-1)(D-2)\frac{\dot{a}^2}{a^2},\label{eq:torsiond}\\
B_{11}=B_{22}=....=B_{(D-1)(D-1)}&=&-2(D-2)^2(D-1)\dot{a}^2\left(\frac{\ddot{a}}{a}-\frac{\dot{a}^2}{a^2}\right),\label{eq:btensor}\\
G_{00}&=&\frac{1}{2}(D-1)(D-2)\frac{\dot{a}^2}{a^2},\label{eq:einsteintensor0}\\
G_{11}=G_{22}=.....=G_{(D-1)(D-1)}&=&-\left[(D-2)a\ddot{a}+\frac{1}{2}(D-2)(D-3)\dot{a}^2\right],\label{eq:einsteintensor1}\\
f_T(T)&=&\left(1+\frac{2T}{D\lambda}\right)^{\frac{D-2}{2}},\label{eq:ft}\\
f_{TT}(T)&=&\left(\frac{D-2}{D\lambda}\right)\left(1+\frac{2T}{D\lambda}\right)^{\frac{D-4}{2}}.\label{eq:ftt}
\end{eqnarray}
It is straight forward to write down the field equations using the expressions (\ref{eq:torsiond}-\ref{eq:ftt}) in Eq.~(\ref{eq:f(t)_covariant}). There are two field equations, one for $\rho$ from $\Theta_{\mu\nu}$ ($\rho$-equation) and another for $p$ ($p$-equation). Further, these two field equations lead to the conservation equation
\begin{equation}
\dot{\rho}+(D-1)H(p+\rho)=0
\label{eq:conservation}
\end{equation}  
where $H=\dot{a}/a$ (the Hubble function). If we assume an equation of state: $p=\omega\rho$, the conservation equation leads to a relation between energy density ($\rho(t)$) and scale factor ($a(t)$) given as $\rho\propto a^{-(D-1)(\omega+1)}$.

\subsection{$3+1$ dimensional cosmological solutions}
\begin{figure}[h]
\begin{center}
\mbox{\epsfig{file=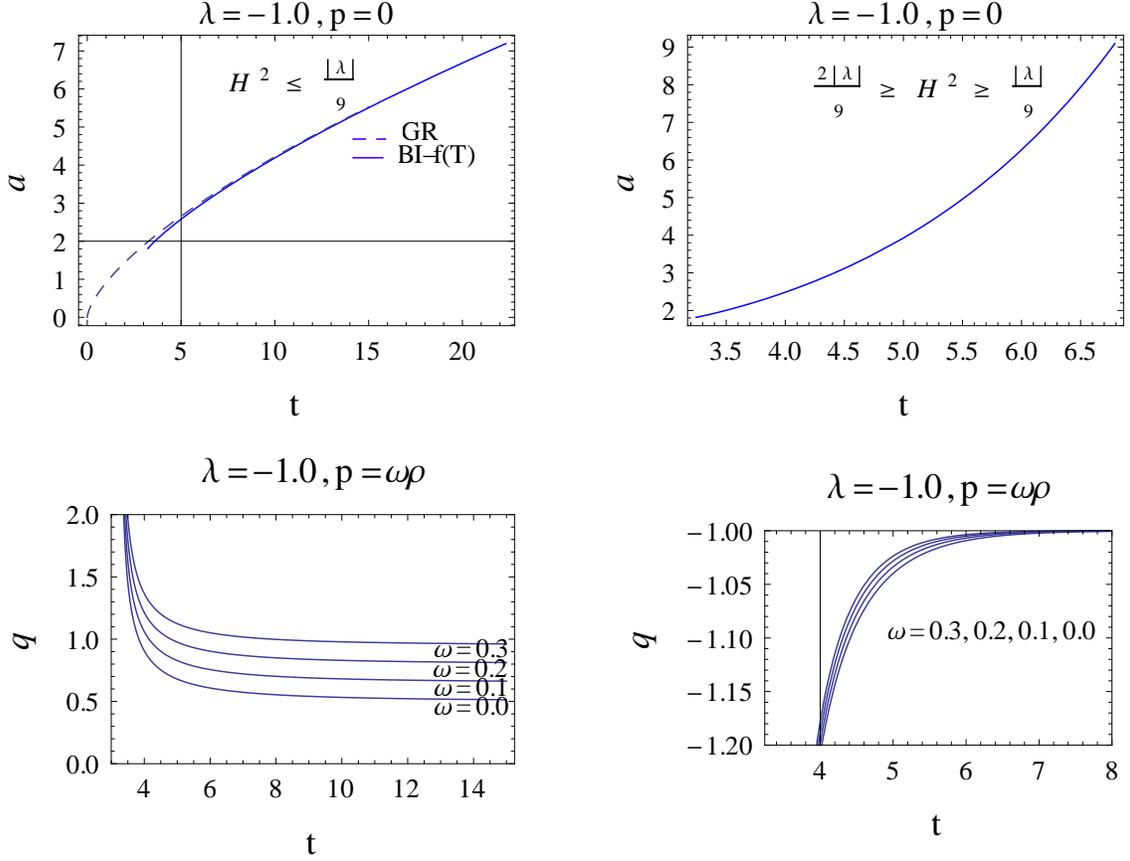,width=6in,angle=360}}
\end{center}
\caption{(Top  panel) Plot of scale factor $a(t)$ for $\lambda=-1.0$, $p=0 $ in $3+1$ cosmology and  comparison with its counterpart in GR. (Bottom panel) plot of deceleration parameter for different equations of state.In the solutions, $C_2=1$, $k_2=0$ and $8\pi G =1$.  }
\label{fig:tebicosmo4lmbdaneg}
\end{figure}

\noindent For the $D=4$ case, i.e. $3+1$ dimensional cosmological model, the $\rho$-equation and $p$-equation are
\begin{eqnarray}
3\frac{\dot{a}^2}{a^2}&=&8\pi G \rho -\frac{27}{2\lambda}\frac{\dot{a}^4}{a^4}\label{eq:rhoequation4}\\
-2\frac{\ddot{a}}{a}-\frac{\dot{a}^2}{a^2}&=&8\pi G p +\frac{9}{2\lambda}\frac{\dot{a}^2}{a^2}\left(4\frac{\ddot{a}}{a}-\frac{\dot{a}^2}{a^2}\right)\label{eq:pequation4}
\end{eqnarray} 
Note the changes in Eqs.~(\ref{eq:rhoequation4}), (\ref{eq:pequation4}) in the R.H.S. of the equations. The L.H.S. is the same as in GR. The modifications may be treated as geometrical contributions to the effective energy density and effective pressure.  Using the equation of state $p=\omega \rho$, $H=\frac{\dot{a}}{a}$, $\frac{\ddot{a}}{a}=\dot{H}+H^2$ and combining the $\rho$-equation and the $p$-equation we get
\begin{equation}
\frac{\left(\frac{1}{H^2}+\frac{9}{\lambda}\right)\dot{H}}{\left(\frac{2}{H^2}+\frac{9}{\lambda}\right)H^2}=-\frac{3(\omega +1)}{4}\label{eq:hequation}
\end{equation}
For $\lambda>0$, we have an analytical solution
\begin{equation}
\frac{1}{H}+\frac{3}{\sqrt{2\lambda}}\tan^{-1}\left(\frac{\sqrt{2\lambda}}{3H}\right)=\frac{3}{2}(\omega +1)t + k_1
\label{eq:solheqlmbdpos}
\end{equation}
where $k_1$ is an arbitrary constant. The
conservation equation leads to a relation $\rho=\frac{C_2}{a^{3(\omega +1)}}$ ($C_2$ is a constant). Using this relation in Eq.~(\ref{eq:rhoequation4}), we get 
\begin{eqnarray}
H^2&=&\frac{\lambda}{9}\left[\sqrt{1+\frac{48\pi G C_2}{\lambda a^{3(\omega +1)}}}-1\right]\label{eq:hsqure_lambdapos}\\
\mbox{or,}\quad~ a&=& \left[\frac{16\pi G C_2}{6H^2+\frac{27}{\lambda}H^4}\right]^{\frac{1}{3(\omega +1)}}.
\label{eq:a(h)} 
\end{eqnarray}
Eq.~(\ref{eq:solheqlmbdpos}) and Eq.~(\ref{eq:a(h)}) can give an analytical expression for $a(t)$, where $H$, the Hubble function, is independent and we have $0<H<\infty$. We note that there is no upper limit of $H$ and as a consequence $a$ has no lower limit for $\lambda >0$. Thus we have a singular solution. On the other hand, when $t\rightarrow \infty$, then $a\rightarrow \infty$, $H^2\approx \frac{8\pi G C_2}{3a^{3(\omega +1)}}$ and $a(t)\sim t^{\frac{2}{3(\omega +1)}}$ (GR limit). The deceleration parameter ($q=-\frac{\ddot{a}}{aH^2}$) expressed as a function of $H$ is:
\begin{equation}
q=\frac{3(\omega +1)}{4}\left(\frac{\frac{2}{H^2}+\frac{9}{\lambda}}{\frac{1}{H^2}+\frac{9}{\lambda}}\right)-1
\label{eq:q(h)}
\end{equation}  
For $a\rightarrow 0$, $H\rightarrow \infty$, $q\rightarrow \frac{3(\omega +1)}{4}-1$. Note that if $\omega \leq 1/3$ then $q(H\rightarrow \infty)\leq 0$ implying accelerated expansion of the early universe. For $a\rightarrow 0$, $a(t)\sim (t-t_0)^{\frac{4}{3(\omega +1)}}$ which shows that $a$ becomes zero in the finite past and the singularity is a big-bang singularity.
\begin{figure}[h]
\begin{center}
\mbox{\epsfig{file=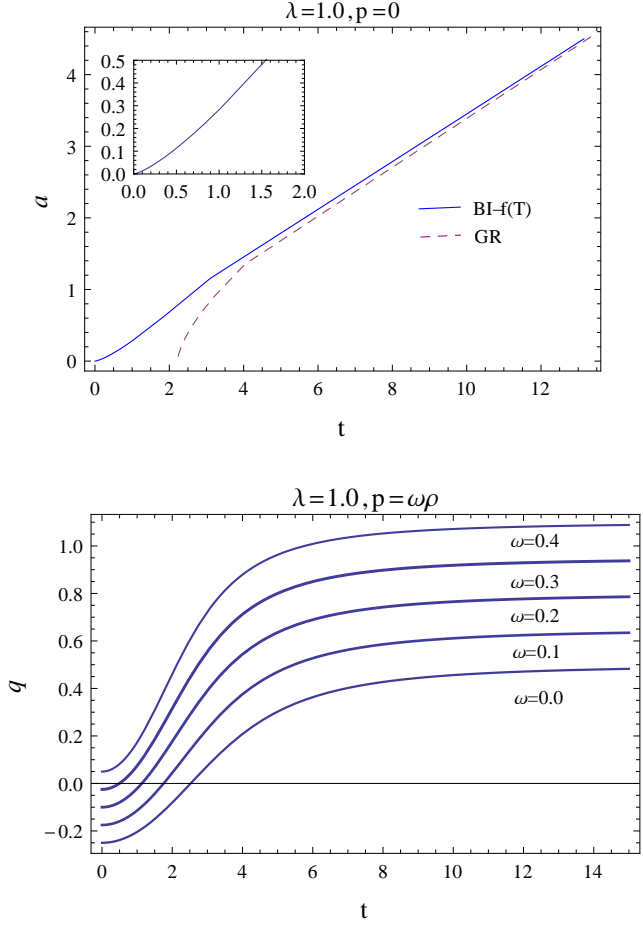,width=4in,angle=360}}
\end{center}
\caption{(Top panel) Plot of scale factor $a$ for $\lambda=1.0$, $p=0 $ in $3+1$ cosmology and  comparison with its counterpart in GR. (Bottom panel) Plot of deceleration parameter ($q$) for different equations of state. In the solutions, $C_2=1$, $k_1=0$ and $8\pi G =1$. }
\label{fig:tebicosmo4lmbdapos}
\end{figure}
In Fig.~\ref{fig:tebicosmo4lmbdapos}, on the top panel, the plot of the scale factor ($a$) shows an accelerated expansion of an early dust-filled-universe (clearly shown in the inset figure) and at late times, the plot merges with GR-solution, as expected. The bottom panel shows the plots of the deceleration parameter for different equations of state. We note that, for $\omega \leq 1/3$, the universe has a phase transition from accelerated expansion state to a decelerated expansion state in a finite future time. \\ 
\noindent Let us now turn to $\lambda<0$. Eq.~(\ref{eq:hequation}) becomes
\begin{equation}
\frac{\left(\frac{1}{H^2}-\frac{9}{\vert\lambda \vert}\right)\dot{H}}{\left(\frac{2}{H^2}-\frac{9}{\vert\lambda\vert}\right)H^2}=-\frac{3(\omega +1)}{4}\label{eq:hequation_lmbdneg}
\end{equation}
This equation has a solution:
\begin{equation}
\frac{1}{H}-\frac{3}{2\sqrt{2\vert \lambda \vert}}\log \left(\frac{\frac{\sqrt{2\vert \lambda \vert}}{3H}-1}{\frac{\sqrt{2\vert \lambda \vert}}{3H}+1}\right)=\frac{3(\omega +1)}{2}t + k_2
\label{eq:solhequation_lmbdneg}
\end{equation}
where $k_2$ is an arbitrary constant.
Now from conservation equation and $\rho$-equation we get
\begin{equation}
H^2=\left(\frac{\dot{a}}{a}\right)^2=\frac{\vert \lambda \vert}{9}\left[1\pm \sqrt{1-\frac{48\pi G C_2}{\vert \lambda \vert a^{3(\omega +1)}}} \right]
\label{eq:friedmannequation}
\end{equation}
The `$\pm$' sign in the equation is to be noted carefully since it indicates two different solutions. For both the solutions, there is a minimum scale factor $a_B=\left(\frac{\vert \lambda \vert}{48\pi G C_2}\right)^{-\frac{1}{3(\omega +1)}}$, where $\rho$ is maximum and $H^2$ has the value $H^2(a_B)=\frac{\vert \lambda \vert}{9}$. For large $a$ ($a\rightarrow \infty$), $H^2\rightarrow 0$ or $\frac{2\vert \lambda \vert}{9}$. Equation~(\ref{eq:solhequation_lmbdneg}) and equation~(\ref{eq:friedmannequation}) with negative sign in the second term in the square bracket on the R.H.S. leads to a
scale factor $a(t)$ representing decelerated expansion  of universe. At late times this solution converges to a GR solution. The positive sign of the second term in the square bracket on R.H.S. of Eq.~(\ref{eq:friedmannequation}) leads to a solution representing accelerated expansion. At late times, we have de Sitter expansion phase $\left(a(t)\sim e^{\frac{\sqrt{2\vert \lambda \vert}}{3}t}\right)$. Both these solutions can be obtained from
\begin{eqnarray}
a&=&\left[\frac{8\pi G C_2}{3H^2\left(1-\frac{9H^2}{2\vert \lambda \vert}\right)}\right]^{\frac{1}{3(\omega +1)}}\\
\mbox{and,}\quad~ t&=&\frac{2}{3(\omega +1)}\left[\frac{1}{H}-\frac{3}{2\sqrt{2\vert \lambda \vert}}\log \left(\frac{\frac{\sqrt{2\vert \lambda \vert}}{3H}-1}{\frac{\sqrt{2\vert \lambda \vert}}{3H}+1}\right)-k_2\right].
\end{eqnarray}
For the first type of solution $H^2\leq \frac{\vert \lambda \vert}{9}$ and for the other $\frac{2\vert \lambda \vert}{9}\geq H^2 \geq \frac{\vert \lambda \vert }{9}$. The expression for the deceleration parameter ($q$) as a function of ($H$) becomes
\begin{equation}
q=\frac{3(\omega +1)}{4}\left(\frac{\frac{2}{H^2}-\frac{9}{\vert \lambda \vert}}{\frac{1}{H^2}-\frac{9}{\vert \lambda \vert}} \right)-1
\label{eq:q(h)_lmbdneg}
\end{equation}
When $H^2\leq \frac{\vert \lambda \vert}{9}$, $q>0$ and for $H\rightarrow 0$, $q(H\rightarrow 0)\approx \frac{3\omega +1}{2}$ (GR limit). On the other hand, when $\frac{2\vert \lambda \vert}{9}\geq H^2 \geq \frac{\vert \lambda \vert}{9}$, $q<0$ and when $H^2\rightarrow \frac{2\vert \lambda \vert}{9}$, $q\rightarrow -1$ (de Sitter expansion stage). Fig.~\ref{fig:tebicosmo4lmbdaneg} also demonstrates these characteristics of the solutions.

At the minimum value of the scale factor ($a_B$), $\frac{\ddot{a}}{a}$ diverges which is evident from Eq.~(\ref{eq:hequation_lmbdneg}). Hence the {\em Ricci scalar} ($R$) also diverges at minimum scale factor, although energy density and pressure remain finite. Thus we do not have a big-bang singularity where the scale factor becomes zero and the {\em Ricci scalar} as well as all physical quantities like $\rho$, $p$ becomes infinite.  

\subsection{Observational test of the theory from fitting the Supernova data}

It is known that the $f(T)$ gravity has been studied with reference to observational cosmology \cite{cosmography,noether_f(T)}.
Now we test the viability of the theory (and the solutions discussed above) with the cosmological observations. We constrain the model parameters from fitting of the Supernova data and using the fitted parameter values, we estimate the properties of the universe and compare those with the known values-- such as the age of the universe, the value of the deceleration parameter etc. To fit the Supernova data with our model, we follow the method used in \cite{supernova_fit}, wherein the authors have studied the expansion history of the universe upto a redshift $z=1.75$ using the 194 Type Ia supernovae (SNe Ia) data published in \cite{sneia_torny,sneia_barris}. We define the Hubble free luminosity distance ($d_L$) using $D_L=cH_0^{-1}d_L$, where $D_L$ is the luminosity distance, $H_0$ is present day observed value of the Hubble parameter and $c$ is the speed of light. So the expression of the Hubble free luminosity distance is given by
\begin{equation}
d_L(z)=(1+z)\int^{z}_{0}dz' \frac{H_0}{H(z')} 
\label{eq:d_L}
\end{equation}    
The observational dataset consists of apparent magnitudes $m_i(z_i)$ and redshifts $z_i$ with their corresponding errors $\delta m_i$ and $\delta z_i$. Each apparent magnitude is related to the corresponding luminosity distance $D_L$ of the SNe Ia by 
\begin{equation}
m(z)= M + 5log_{10}\left[\frac{D_L(z)}{Mpc} \right]+25
\label{eq:apparent_m}
\end{equation}
where $M$ is the absolute magnitude which is assumed to be constant for standard candles like  SNe Ia.
Using the definition of the Hubble free luminosity distance ($d_L$), Eq.~(\ref{eq:apparent_m}) can be rewritten as 
\begin{equation}
m(z)=\bar{M}(M,H_0)+5log_{10}(d_L(z))
\end{equation}
where $\bar{M}$ is the magnitude zero point offset expressed as
\begin{equation}
\bar{M}=M+5log_{10}\left[\frac{c/H_0}{1 Mpc}\right]+25
\end{equation}
The observed $m_i(z_i)$ can be translated to $d^{obs}_L(z_i)$ for the best fit value of $\bar{M}_{obs}$ obtained from nearby SNe Ia \cite{supernova_fit}. For a given model $H(z;a_1,a_2,...,a_n)$, one can also theoretically predict the $d^{th}_L(z)$ using the Eq.~(\ref{eq:d_L}). The best fit values of the model parameters ($a_1,a_2,....,a_n$) are estimated by minimizing the $\chi^2(a_1,a_2,....,a_n)$ which, in this case, is given by \cite{supernova_fit}
\begin{equation}
\chi^2(a_1,a_2,....,a_n)=\sum_{i=1}^{194}\frac{\left(log_{10}d_L^{obs}(z_i)-log_{10}d_L^{th}(z_i)\right)^2}{\left(\sigma_{log_{10}d_L(z_i)}\right)^2+\left(\frac{\partial log_{10}d_L(z_i)}{\partial z_i}\sigma_{z_i}\right)^2}
\end{equation} 
where $\sigma_z$ is $1\sigma$ redshift uncertainty of the data and $\sigma_{log_{10}d_L(z_i)}$ is $1\sigma$ error of $log_{10}d_L^{obs}(z_i)$. We use the same table of data which was used by the authors in \cite{supernova_fit} and it can be downloaded from \cite{supernova_data}. Each row of the table of data contains redshift $z$, $log_{10}(cd^{obs}_L(z))$ and the corresponding error $\sigma_{log_{10}d_L}$. The error in redshift $\sigma_z$ is estimated from uncertainty due to peculiar velocities, $\Delta v=\Delta (cz)=500 \, km/s$, i.e. $\sigma_z=\Delta z= (500 \, km/s)/c$.\\

\begin{figure}[h]
 \begin{center}
 \begin{tabular}{cc}
        \resizebox{100mm}{!}{\includegraphics{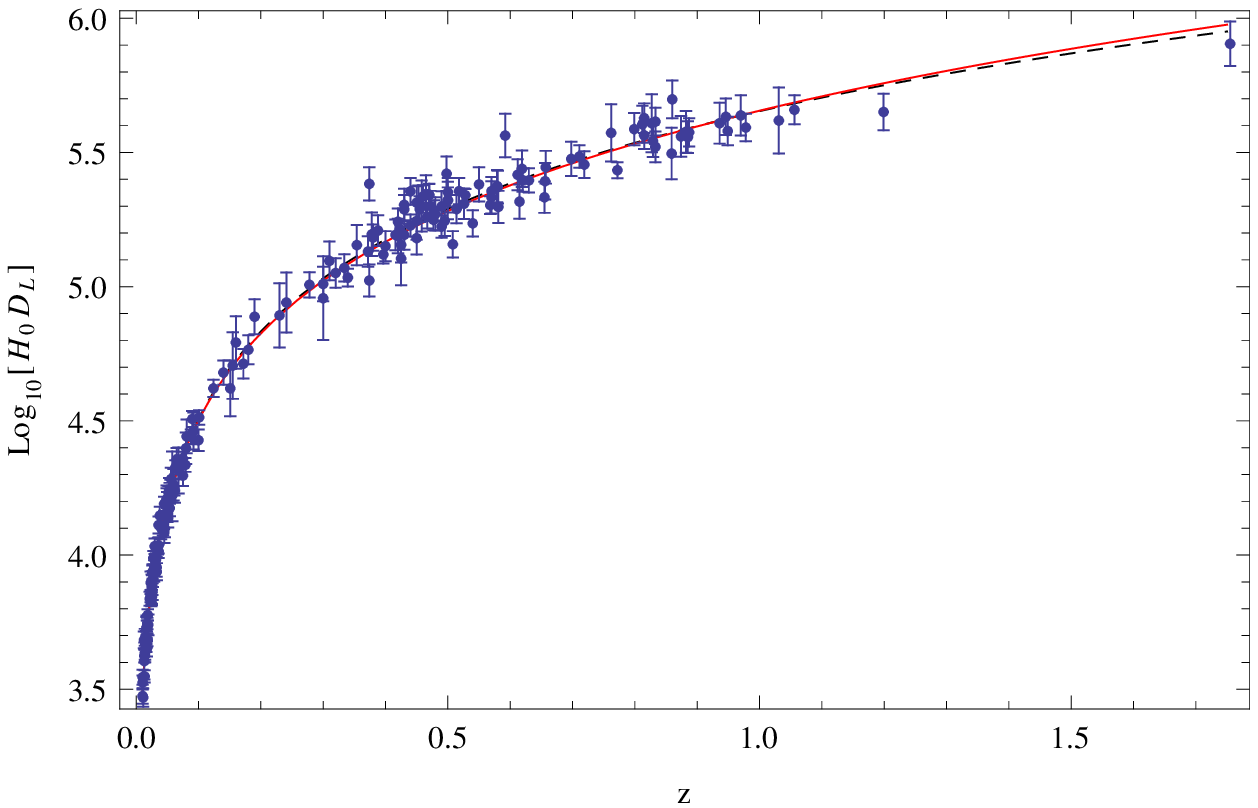}}&
        \resizebox{65mm}{!}{\includegraphics{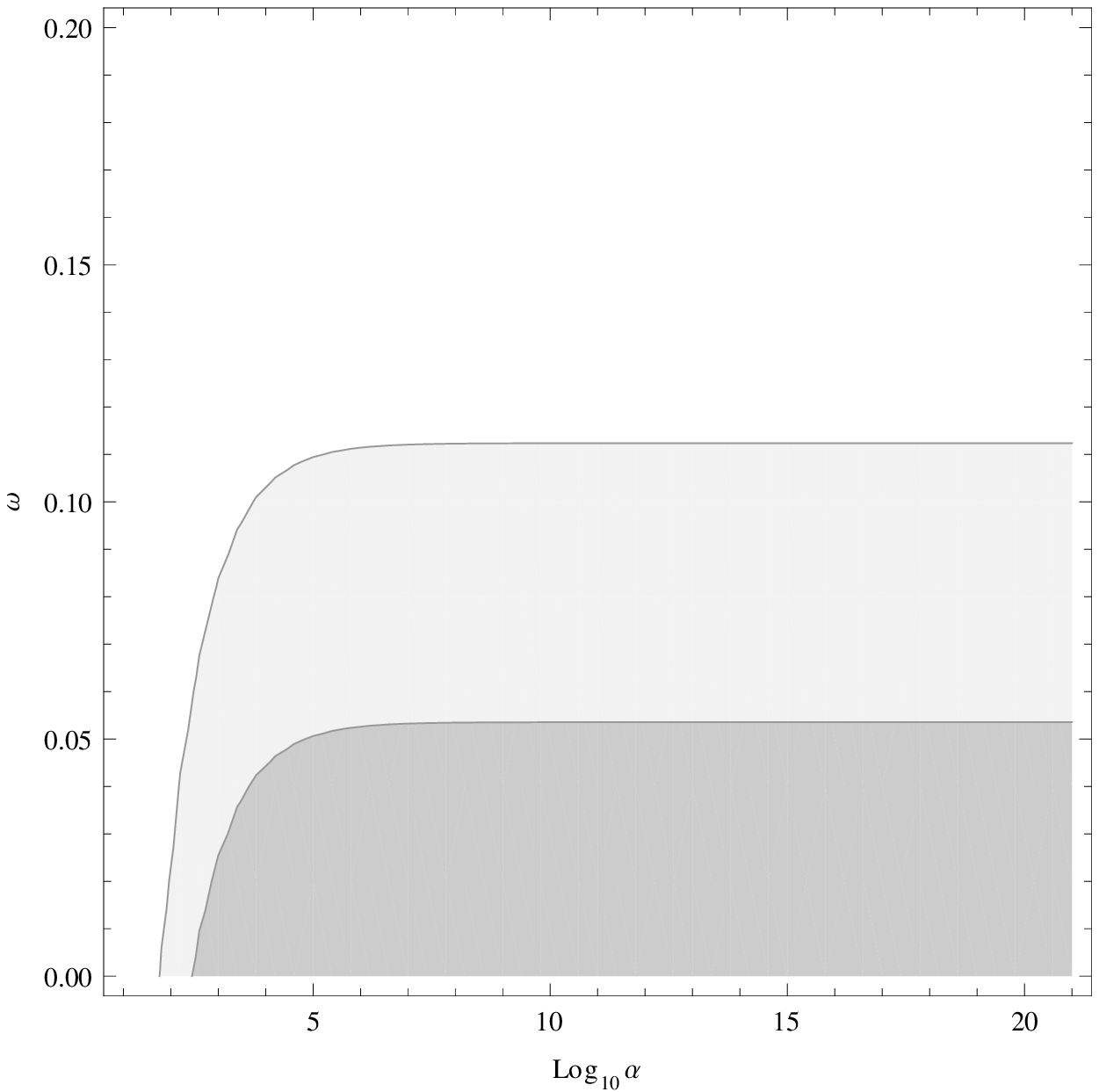}}
        
 \end{tabular}
 \caption{In the left panel, observed 194 SNe Ia Hubble free luminosity distance along with the fitted curve (solid red line), using the model [Eq.~(\ref{eq:model1})], is shown and it is also compared with the $\Lambda$CDM model (black dashed line)\cite{supernova_fit}. In the right panel, $1\sigma$ (inner, thick gray-shaded region) and $2\sigma$ (including inner shaded region and the outer, light gray-shaded region) error plots are shown in $2$-D parameter space ($\omega-log_{10}\alpha$). }
 \label{fig:model1}
 \end{center}
 \end{figure}

 Now we apply this method to our model. First, we bring back $c$ in the field equations--earlier, it was assumed that $c=1$. Then, the torsion scalar becomes $T=\frac{6H^2}{c^2}$ and in all the equations and solutions, following replacements should be made: $\lambda\rightarrow \lambda c^2$ and $p\rightarrow p/c^2$. Here, we note that both $\lambda$ and the torsion scalar ($T$) have the dimension of $1/(distance)^2$. Similarly, the equation of state should be: $p=\omega \rho c^2$. In our analysis, we stress upon the fact that there are contributions from spacetime geometry like the energy density and pressure in the field equations and these are the terms involving $\lambda$. Here, we rewrite the $\rho$-equation [Eq.~(\ref{eq:rhoequation4})] in the following way:
\begin{eqnarray}
3\left(\frac{\dot{a}}{a}\right)^2&=&8\pi G \rho -\frac{27}{2\lambda c^2}\left(\frac{\dot{a}}{a}\right)^4
=8\pi G \rho_{eff}\nonumber \, , \\
\rho_{eff}&=& \rho - \frac{27}{16\pi G \lambda c^2}\left(\frac{\dot{a}}{a}\right)^4
\label{eq:rho_eff}
\end{eqnarray}
where, $\rho_{eff}$ is the ``effective energy density" which consists of $\rho$, the energy density coming from the stress-energy tensor, and the second term, coming from spacetime geometry. Then the present day (redshift $z=0$) value of the effective energy density is same as the present day value of  the so-called critical density, i.e. $\rho_{eff}(z=0)=\rho_{c0}=3H_0^2/{8\pi G}$. Now if we assume the physical matter, filling the universe, to be mostly the dust-like, then $\rho=\Omega_{m0}\rho_{c0}(1+z)^3$. Then, from the Eq.~(\ref{eq:rho_eff}), we get that
\begin{equation}
\frac{12\pi G \rho_{c0}}{\lambda c^2}=\Omega_{m0}-1
\label{eq:omega_m0}
\end{equation}    
So, if $\lambda$ is a finite positive value then, from Eq.~(\ref{eq:omega_m0}), it is evident that $\Omega_{m0}>1$. But, from the WMAP7 \cite{wmap7} determination of the physical matter density, it is known that $\Omega_{m0}\approx 0.3$. So, this indicates that $\lambda$ is to be a negative value which further leads to a solution for the scale factor $a(t)$ having a minimum value $a_B$ [Eq.~(\ref{eq:friedmannequation})] or, consequently, a maximum finite value of redshift ($z_{\infty}$). In this case, $z_{\infty}$ turns out to be $0.06$ which is absurd. At least, from CMBR spectrum, we may expect $z_{\infty}>1000$. However, we let $\Omega_{m0}$ be a free parameter and fit the Supernova data with the following model (derived from Eq.~\ref{eq:hsqure_lambdapos} ):
\begin{equation}
d_L(z; \alpha,\omega)=(1+z)\int^z_0\left[\frac{\sqrt{1+\alpha}-1}{\sqrt{1+\alpha (1+z')^{3(\omega +1)}}-1} \right]^{\frac{1}{2}}dz'
\label{eq:model1}
\end{equation}    
where, $\omega$ is also a free parameter, but with a constraint $\omega > 0$. Here, $\alpha=\Omega_{m0}\frac{48\pi G\rho_{c0}}{\lambda c^2}$. The best fit value of the parameters are $\alpha=2.15\times 10^8$ and $\omega = 0$ ($i.e.$ dust), with $\chi^2_{min}/{d.o.f.}= 1.06$. Fig.~\ref{fig:model1} shows that the Supernova data fitting, with this model, is very close to $\Lambda$CDM model \cite{supernova_fit}. The best fit parameter-values have been estimated and the figure has been plotted by use and modification of the {\em Mathematica} code available in \cite{supernova_data}. We use a prior value $H_0=70\, km/s/Mpc$ and estimate $\lambda=3.342\times 10^{-11}\, Mpc^{-2}$. 

Using the estimated parameter values, we find that $\Omega_{m0}=7331.94$, which indicates more than twenty thousand times greater value of the physical matter density than what is estimated from WMAP7. The reason behind it is that though the geometrical contribution to the  ``effective pressure" favors the acceleration, it does so at the price of a negative energy density contribution in the total ``effective energy density" and it is so high in magnitude that, in compensation, the physical matter density  also becomes very high. Moreover, we estimate the age of the universe as $t_0\simeq 18.67$ Billion years, where as its presently accepted value is close to 14 Billion years. The present value of the deceleration parameter is also estimated and it is found to be $q_0\simeq -0.25$; but from the  cosmological observations, we know that it should be close to a value of $-0.64$. So, it is clear that this model though fits the Supernova data well, is not in good agreement with expected model independent properties of the universe. However, this model is better than SCDM model (the model in GR with only dust as physical matter of the universe but without cosmological constant)\cite{supernova_fit} which does not provide acceleration at all. 
A similar analysis tells us that the models with negative-$\lambda$ solutions do not fit good to the data.\\ 

Now, we turn to investigate that if we can have a better model by the addition of an {\em extra} constituent along with the dust in $\rho$. We use a prior value $\Omega_{m0}=0.3$ for the dust and assume that $\rho=0.3\rho_{c0}(1+z)^3+\Omega_{ext}\rho_{c0}(1+z)^{3(\omega_{e}+1)}$. We find that $\Omega_{ext}-\Omega_{\lambda}=0.7$, where $\Omega_{\lambda}=\frac{12\pi G \rho_{c0}}{\lambda c^2}$, so that the remaining part, except the dust, in the critical density $\rho_{c}$, equals to the sum of the extra constituent and the geometrical contribution. Now Eq.~(\ref{eq:model1}) becomes

\begin{equation}
d_L(z; \Omega_{\lambda},\omega_{e})=(1+z)\int^z_0\left[\frac{2\Omega_{\lambda}}{\sqrt{1+4\Omega_{\lambda}\lbrace 0.3(1+z')^3+(0.7+\Omega_{\lambda})(1+z')^{3(\omega_e +1)}\rbrace}-1} \right]^{\frac{1}{2}}dz'
\label{eq:model2}
\end{equation}

The best fit value of the parameters in this model are estimated and found to be $\Omega_\lambda=5.992\times 10^{-9}$ and $\omega_{e}=-0.923$, with $\chi^2_{min}/{d.o.f.}=1.04$. In this model, we found $\lambda=41.0\, Mpc^{-2}$, the age of the universe $t_0=13.33$ Billion years and the present day value of the deceleration parameter $q_0=-0.47$.

\begin{figure}[h]
 \begin{center}
 \begin{tabular}{cc}
        \resizebox{90mm}{!}{\includegraphics{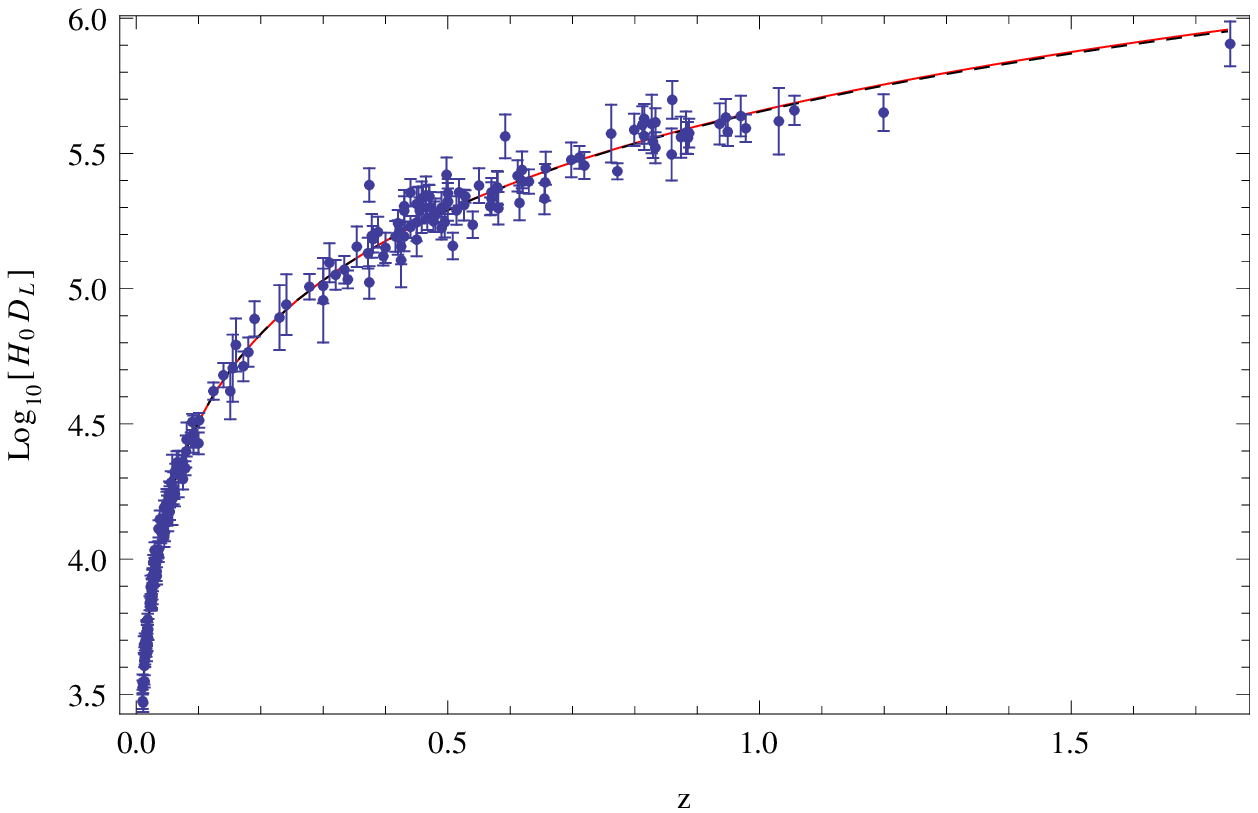}}&
        \resizebox{65mm}{!}{\includegraphics{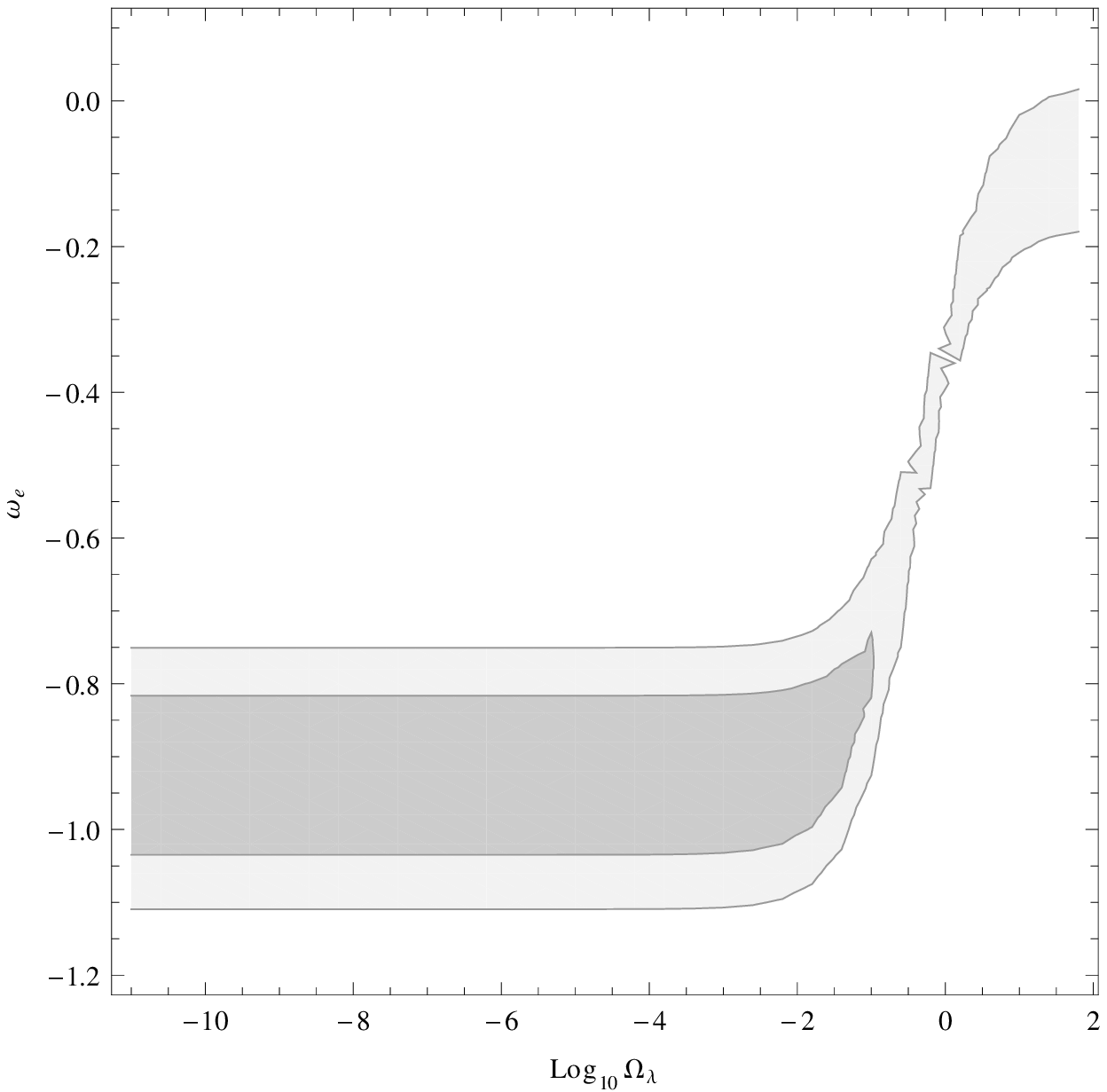}}\\
        \resizebox{90mm}{!}{\includegraphics{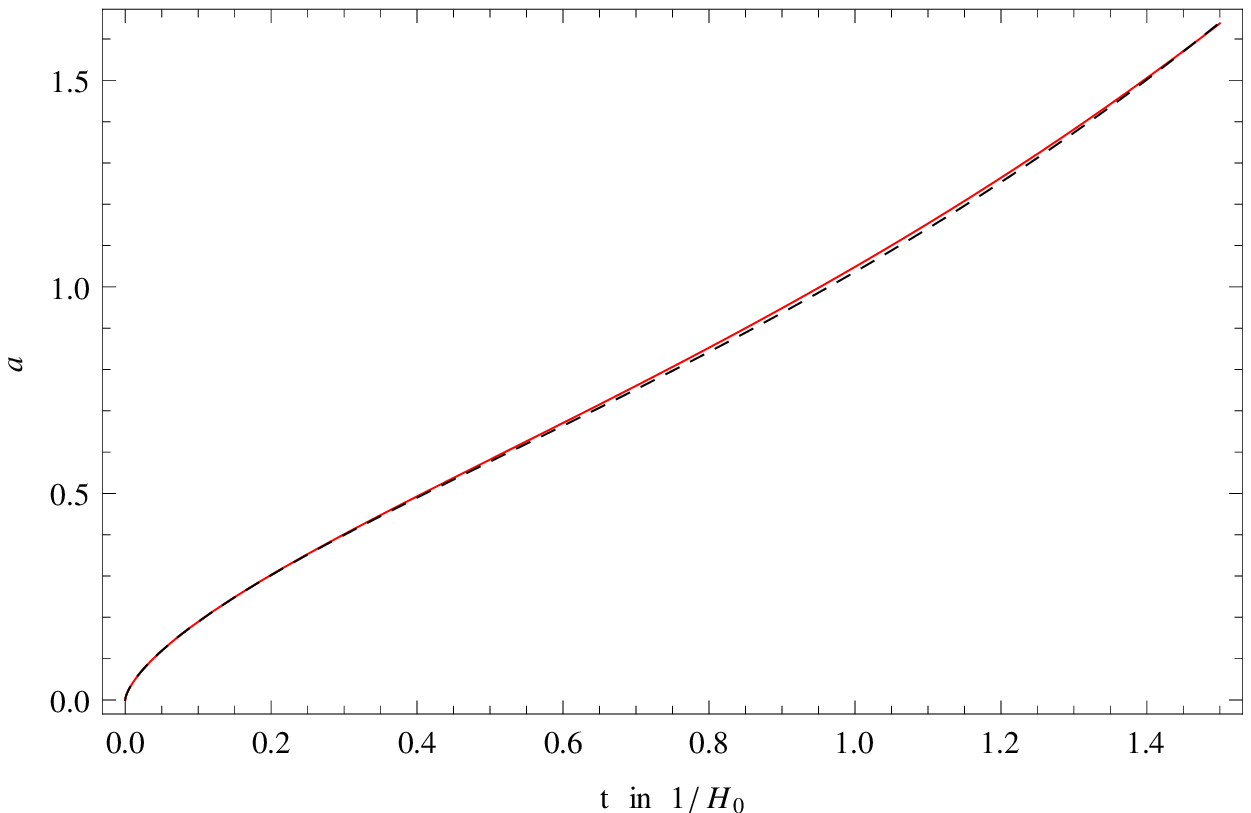}}&
        \resizebox{65mm}{!}{\includegraphics{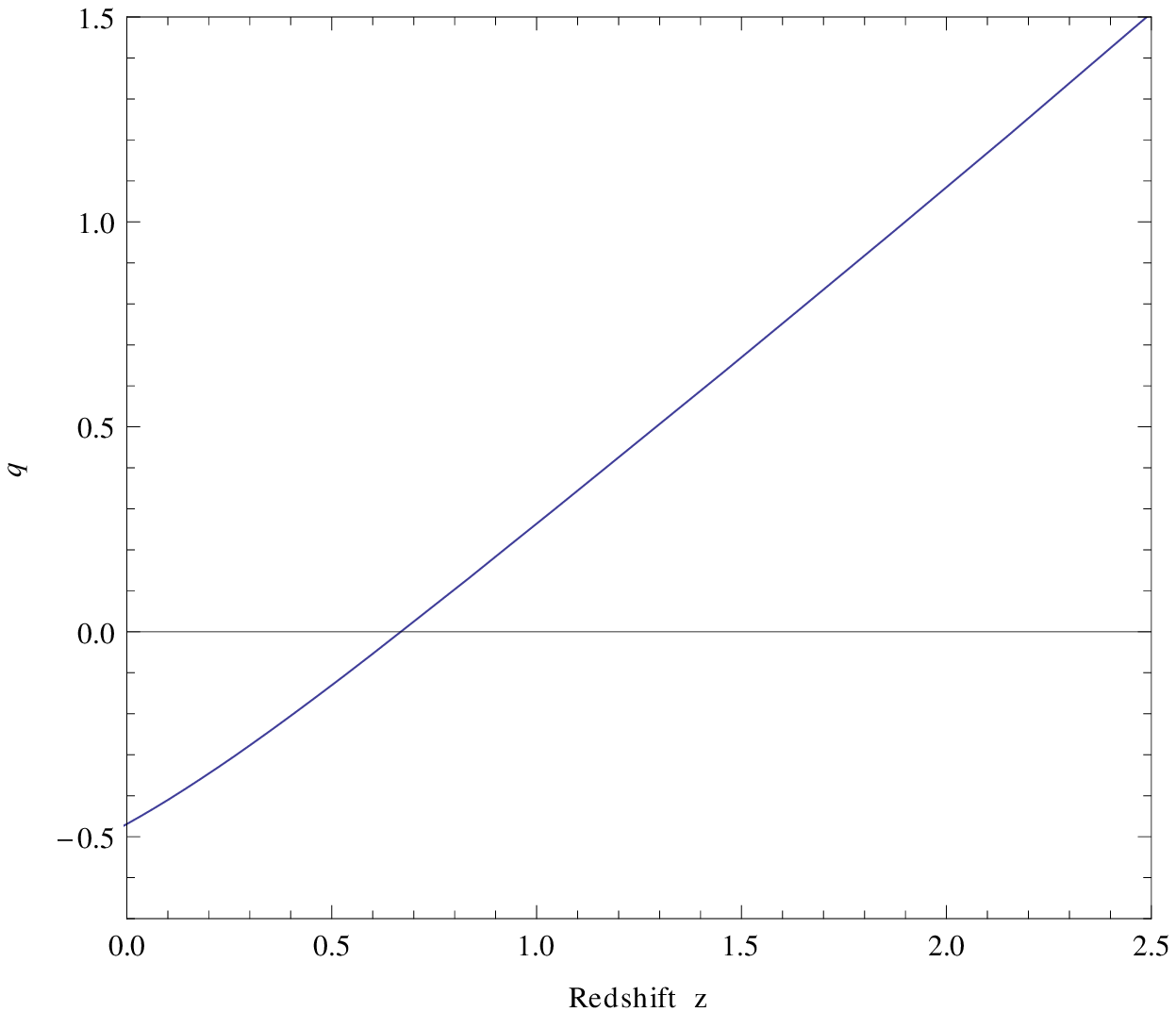}} 
        
 \end{tabular}
 \caption{In the top-left panel, observed 194 SNe Ia Hubble free luminosity distance along with the fitted curve (solid red line), using the model [Eq.~(\ref{eq:model2})], is shown and it is also compared with the $\Lambda$CDM model (black dashed line)\cite{supernova_fit}. In the top-right panel, $1\sigma$ and $2\sigma$ error plots are shown in $2$-D parameter space ($\omega_e-log_{10}\Omega_{\lambda}$). In the bottom-left panel, the scale factor has been plotted as a function of time using the best fit parameter values (the solid red line) and it is compared with the same for $\Lambda$CDM model (the black dashed line); the unit of time used is $1/H0\approx 14\, GYr$. In the bottom-right panel, the deceleration parameter ($q$) is plotted as a function of redshift parameter ($z$).   }
 \label{fig:model2}
 \end{center}
 \end{figure}

 This model is very close to $\Lambda$CDM model in GR, since all the estimated properties of the universe and the scale factor evolution matches very closely with $\Lambda$CDM model (see Fig.~\ref{fig:model2}). Actually, the ratio of the present day value of the torsion scalar ($T_0$) and $\lambda$ is so small ($T_0/\lambda=8.0\times 10^{-9}$) that we can say that, in this model, the law of gravity is mostly governed by GR. Moreover, in this model, we use an extra constituent with $\omega_e=-0.923$ (quite similar to GR), and therefore, it does not give an explanation of the dark energy problem. \\
We further work with the models like $\omega_{e}(z)=\omega_0+\omega_1 z$ and $\omega_{e}(z)=\omega_0+\omega_1z/(1+z)$, with the hope that $\omega_{e}(z)>0$ for all $z$. But, we do not see much qualitative difference from GR. Also, such models may have further problems like future singularities.

\section{Conclusions}
\label{sec:conclusion}
In this article we have examined the cosmology resulting from a new Born-Infeld like $f(T)$ theory of gravity. We have considered a spatially flat universe  driven by a perfect fluid with non negative pressure.

The cosmological solutions presented here possess the following features. If $\lambda>0$ the solutions are singular of big-bang type but may have an early accelerated expansion phase though not inflationary. If $\lambda<0$ the solutions are still singular but not of big-bang type, rather these are ``softened" as there is a non-zero minimum scale factor and finite maximum values of energy density and pressure. The curvature scalar diverges and hence, such singularity is a purely geometrical feature.  This type of singularity has been reported earlier in different contexts \cite{typeivsingularity}. In \cite{suddensingularity} it is shown that such singularity also occurs in a different model of the Born-Infeld type (\cite{fiorini}) cosmology  and the authors call it as the ``Sudden singularity".

Interestingly, if $\lambda <0$, there may be an eternal accelerated expansion of the universe with a de Sitter expansion phase at late times. This is an intriguing result because we have assumed the universe filled with only ordinary matter ($p=\omega \rho$, $\omega>0$). Thus the late time acceleration of the universe is a natural consequence of the theory and it boosts the belief that such a  modification to GR can give a plausible alternative explanation of dark energy. Particularly this feature is absent in other Born-Infeld type theories like the recently proposed Eddington-inspired Born-Infeld gravity \cite{banados,eibicosmo}.  

It is worth mentioning that, in \cite{ferraro}, the authors have considered a relatively straight forward extension of TEGR along the lines of Born-Infeld scalar Lagrangian for electrodynamics \cite{born}. Our results are quite different from the results presented in \cite{ferraro}, where they have shown how the early universe can possess a natural inflationary stage and during this phase the scale factor asymptotically reaches zero value in an infinite past time. The curvature scalar is regular though energy density and pressure diverge.  

One major purpose of such a modification to the theory of gravity, GR, is to explain the observed acceleration of the universe with an extra geometrical term in the field equations, but, without invoking any extra matter constituent of the universe such as the dark energy. Cosmology has been studied well in the $f(T)$-theory of gravity with different chosen forms of $f(T)$. In our article, the form of $f(T)$ is not ad hoc, but, derived from a general Born-Infeld theory of gravity in teleparallel approach. Fitting the Type Ia supernovae data with the cosmological solutions of this Born-Infeld-$f(T)$ theory of gravity throws some light on the plausibility of a geometrical explanation of the dark energy. We see in our first model (Eq.~\ref{eq:model1}) that, indeed, it may be possible. But the problem with this model is that, though it is able to generate the acceleration, the estimated properties of the universe differ widely from currently accepted values. The best model is found to be that where we assume an additional constituent of negative pressure (best fit value of $\omega=-0.923$) along with the dust, in the physical matter density. However, the theory of gravity, considered here, is just one class of Born-Infeld theories of gravity and it may be possible to find more meaningful results in some other variations.
   
Finally we conclude with a few relevant questions. Can we get a model of natural inflationary universe in the same framework ? The answer lies in the fact related to the violation of Lorentz invariance in $f(T)$ theory, which leaves the possibility of finding a new class of tetrads (not a diagonal one) for that purpose. We have seen that if $\lambda <0$ we have two kinds of solutions. Here cosmological fluctuations may be important in the selection of either of these and hence it is worth exploring. Further, we can ask-- can we get new vacuum spherically symmetric static spacetimes different from the black holes in GR ? What is the role of the torsion scalar in such a situation? Finding answers to all of the above questions may give useful information which can shed light further on the viability of the theory .

\section*{Acknowledgments}
The author is grateful to Sayan Kar for his constant inspiration and guidance to carry out this work successfully and also for careful reading of the manuscript. The author acknowledges Rajibul Shaikh for a useful discussion regarding solving an equation in this article. The author also sincerely acknowledges the valuable suggestions of the anonymous referee, which helped in considerable improvement of the article.  

\appendix*

\section{$2+1$ dimensional cosmology}
\noindent In $D=3$ case, the $\rho$-equation and $p$-equation are given as,
\begin{eqnarray}
\sqrt{1+\frac{4}{3\lambda}\frac{\dot{a}^2}{a^2}}\left(\frac{4}{3}\frac{\dot{a}^2}{a^2}-\frac{\lambda}{2}\right)+\frac{\lambda}{2}=8\pi G\rho\label{eq:rhoequation_3}\\
\sqrt{1+\frac{4}{3\lambda}\frac{\dot{a}^2}{a^2}}\left(\frac{\ddot{a}}{a}+\frac{1}{3}\frac{\dot{a}^2}{a^2}-\frac{\lambda}{2}\right)+\frac{\lambda}{2}+\frac{\frac{4}{3\lambda}\frac{\dot{a}^2}{a^2}\left(\frac{\ddot{a}}{a}-\frac{\dot{a}^2}{a^2}\right)}{\sqrt{1+\frac{4}{3\lambda}\frac{\dot{a}^2}{a^2}}}=-8\pi G p\label{eq:pequation_3}
\end{eqnarray}
We now analyze Eqs.(\ref{eq:rhoequation_3}),(\ref{eq:pequation_3}) for the scale factor ($ a(t)$) for both positive and negative values of $\lambda$ with the equation of state $p=\omega \rho$ ( $\rho\propto a^{-2(\omega+1)}$). For a positive value of $\lambda$, the $\rho$-equation (\ref{eq:rhoequation_3}) indicates that as $a\rightarrow \infty$, $\frac{\dot{a}^2}{a^2}\rightarrow 0$ and as $a$ is decreased from $\infty$, $\frac{\dot{a}^2}{a^2}$ increases monotonically without any upper limit, leading to a singularity when the scale factor ($a$) becomes zero. For sufficiently small value of $\frac{\dot{a}^2}{a^2}$, one can expand the L.H.S. of the $\rho$-equation (\ref{eq:rhoequation_3}) upto lowest order: $8\pi G \rho = \frac{\dot{a}^2}{a^2}+\mathcal{O}\left(\frac{\dot{a}^4}{a^4}\right)$, which means that the evolution of the late time universe is governed by GR. But at early times, the evolution is different from GR. To see this, we combine the $\rho$-equation and the $p$-equation to get
\begin{equation}
\frac{\ddot{a}}{a}=H^2-\frac{8\pi G (\omega +1)\rho\sqrt{1+\frac{4H^2}{3\lambda}}}{1+\frac{8H^2}{3\lambda}}
\label{eq:ddota}
\end{equation}  
We solve (\ref{eq:ddota}) numerically and find the behaviour of deceleration parameter ($q$) in this modified theory. We know that, in $2+1$ dimension, for a dust filled universe ($p=0$), in GR, $q=0$. 
\begin{figure}[h]
\begin{center}
\mbox{\epsfig{file=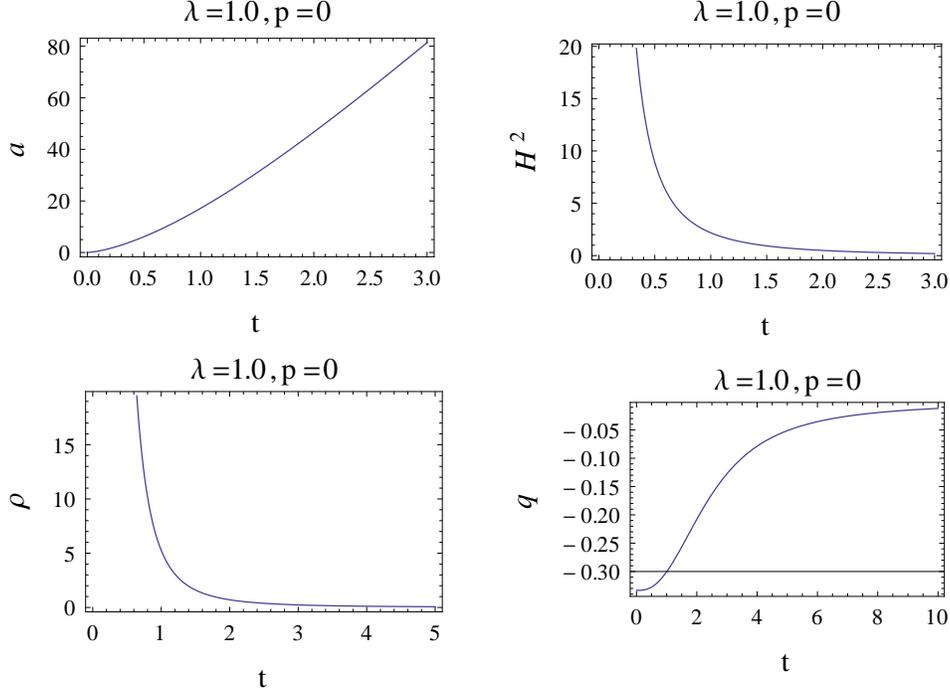,width=5in,angle=360}}
\end{center}
\caption{Plot of scale factor ($a(t)$), Hubble function ($H(t)$), energy density ($\rho(t)$) and deceleration parameter ($q(t)$) for $p=0$, $\lambda=1.0$ in $2+1$ cosmology. In the numerical solutions, the initial conditions are assumed as: $a(t=0)=0.001,\dot{a}(t=0)=1.0, 8\pi G =1 $. }
\label{fig:tebicosmo3lmbdposp0}
\end{figure}
But in Fig.~\ref{fig:tebicosmo3lmbdposp0}, we note that the deceleration parameter ($q$) is always negative, implying an accelerated expansion of the universe, though it approaches a zero value asymptotically at late times. 


We also vary the equation of state parameter ($\omega$) and try to find the behaviour of the deceleration parameter.
\begin{figure}[h]
\begin{center}
\mbox{\epsfig{file=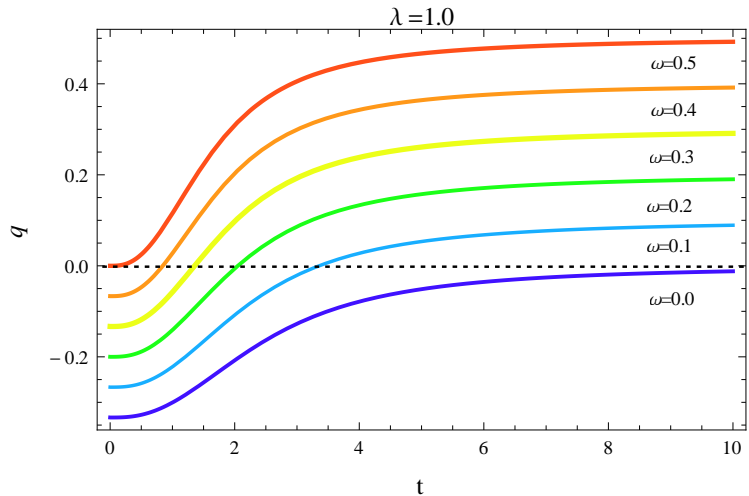,width=4in,angle=360}}
\end{center}
\caption{Plot of deceleration parameter for different $\omega$, for $\lambda=1.0$ in $2+1$ cosmology. }
\label{fig:deceleration3}
\end{figure} 
In Fig.~\ref{fig:deceleration3}, for $0<\omega<1/2$, we note that there is accelerated expansion of 
the early universe but there is a finite future time where the evolution of the universe changes over from accelerated expansion to a decelerated expansion phase. For all $\omega$, the deceleration parameter reaches its GR limit at late times.

\

\noindent Let us now move on to negative-$\lambda$ solutions. The $\rho$-equation (\ref{eq:rhoequation_3}) becomes
\begin{equation}
\sqrt{1-\frac{4}{3\vert\lambda\vert}\frac{\dot{a}^2}{a^2}}\left(\frac{4}{3}\frac{\dot{a}^2}{a^2}+\frac{\vert\lambda\vert}{2}\right)-\frac{\vert\lambda\vert}{2}=8\pi G\rho\label{eq:rhoequation_3_nlmbd}
\end{equation}
From an analysis of the L.H.S. of Eq.~(\ref{eq:rhoequation_3_nlmbd}), we find that there is a minimum scale factor: $a_B=\left[\frac{16\pi G C_1}{\vert\lambda\vert(\sqrt{2}-1)}\right]^{\frac{1}{2(\omega+1)}}$, where $C_1$ is a constant $\left( \rho=\frac{C_1}{a^{2(\omega+1)}}\right)$. At $a_B$, energy density is maximum and its value is $\rho_B=\frac{\vert \lambda\vert(\sqrt{2}-1)}{16\pi G}$. 
A close inspection of Eq.~(\ref{eq:rhoequation_3_nlmbd}) reveals that we have two kinds of solutions: $(i)$ the Hubble function $H=\dot{a}/a$ has a finite maximum value  at the minimum scale factor ($H^2\vert_{a=a_B}=3\vert\lambda\vert/8$) and it decreases monotonically as $a$ increases with time and asymptotically approaches to zero value for large $a$; $(ii)$ as $a$ increases the Hubble function ($H$) also increases from its value at $a_B$ and asymptotically approaches to a maximum value $\sqrt{\frac{3\sqrt{3}\vert\lambda\vert}{8}}$ for $a\rightarrow \infty$. This becomes more evident from Eq.~(\ref{eq:ddota}) which now becomes
\begin{equation}
\frac{\ddot{a}}{a}=H^2+\frac{3\vert\lambda\vert\pi G (\omega +1)\rho\sqrt{1-\frac{4H^2}{3\vert\lambda\vert}}}{H^2-\frac{3\vert \lambda \vert}{8}}
\label{eq:ddota_nlmbd}
\end{equation} 
In the second term of the R.H.S. of Eq.~(\ref{eq:ddota_nlmbd}), the sign of the denominator changes for the two types of solutions and this creates the difference. For one type of solution [$(i)$], $H^2\leq \frac{3\vert \lambda \vert}{8}$ and the second term is always negative. This type of solution describes decelerated expansion of the universe and at late times reduces to the solution in GR ($a\sim t^{\frac{1}{\omega +1}}$). For the other type [$(ii)$-solution], $\frac{3\sqrt{3}\vert \lambda \vert}{8}\geq H^2 \geq \frac{3\vert \lambda \vert}{8}$, and the second term is always positive. This describes an accelerated expansion of the universe for all time and  for very large $a$, the universe has a de Sitter expansion stage $\left(a\sim e^{\sqrt{\frac{3\sqrt{3}\vert \lambda \vert}{8}}t}\right)$. Similar to the $3+1$ case, here also at $a_B$, $\frac{\ddot{a}}{a}$ diverges since denominator of the second term in the R.H.S. of Eq.~(\ref{eq:ddota_nlmbd}) becomes zero. So these solutions are also  singular though the scale factor has a non-zero minimum value and the energy density and pressure have finite maximum value for $\lambda <0$.
 

All these features, for both positive and negative $\lambda$, are similar to what we have seen in $3+1$ cosmology and these are summarized in the Table~\ref{tab:table1}.

\begin{table*}
\caption{\label{tab:table1}Summary of results}
\begin{ruledtabular}
\begin{tabular}{lp{4cm}p{4cm}p{4cm}}

      & \multicolumn{3}{c}{$D=3$}\\

    & singularity & acceleration & GR limit \\
    \hline
    $\lambda >0$    & big-bang & \textrm{for $a\rightarrow 0$ and $\omega < 1/2$}    & \textrm{ for $a\rightarrow \infty$} \\ \hline
     $\lambda < 0$ & \rr \textrm{$a_B\neq 0, \rho(a_B) \neq \infty $, regular $H$, but $\dot{H}(a_B),R(a_B)\rightarrow \infty $} &\textrm{ for all $t$, if $ \frac{3\sqrt{3}\vert \lambda \vert}{8} \geq H^2 \geq \frac{3\vert \lambda \vert}{8} $}; $a(t\rightarrow \infty)\sim e^{\sqrt{\frac{3\sqrt{3}\vert \lambda \vert}{8}}t} $ & \textrm{for $a\rightarrow \infty$, if $H^2\leq \frac{3\vert \lambda \vert}{8} $} \tn
\hline \hline     
  & \multicolumn{3}{c}{$D=4$}\\

    & singularity & acceleration & GR limit \\
    \hline
    $\lambda >0$    & big-bang & \textrm{for $a\rightarrow 0$ and $\omega < 1/3$}    & \textrm{ for $a\rightarrow \infty$} \\ \hline
     $\lambda < 0$ & \rr \textrm{$a_B\neq 0, \rho(a_B) \neq \infty $, regular $H$, but $\dot{H}(a_B),R(a_B)\rightarrow \infty $} &\textrm{ for all $t$, if $ \frac{2\vert \lambda \vert}{9} \geq H^2 \geq \frac{\vert \lambda \vert}{9} $};$a(t\rightarrow \infty)\sim e^{\sqrt{\frac{2\vert \lambda \vert}{9}}t} $ & \textrm{for $a\rightarrow \infty$, if $H^2\leq \frac{\vert \lambda \vert}{9} $} \tn     
  
\end{tabular}
\end{ruledtabular}
\end{table*}

\end{document}